\documentclass{aa}
\input epsf
\begin{document}
\newcommand{\ia}{\mbox{$f(12\mu)/f(25\mu)$}}
\newcommand{\ib}{\mbox{$f(25\mu)/f(60\mu)$}}
\newcommand{\ic}{\mbox{$f(60\mu)/f(100\mu)$}}
\newcommand{\mfir}{\mbox{$m_{\mbox{\scriptsize FIR}}$}}
\newcommand{\lfir}{\mbox{$L_{\mbox{\scriptsize FIR}}$}}
\newcommand{\mb}{\mbox{$m_{\mbox{\scriptsize B}}$}}
\newcommand{\mfirmb}{\mbox{$m_{\rm FIR-B}$}}
\newcommand{\tlu}{\mbox{$T_L(U)$}}
\newcommand{\tlv}{\mbox{$T_L(V)$}}
\newcommand{\msun}{\mbox{M$_{\odot}$}}
\newcommand{\Ssfr}{\mbox{$\Sigma_{\rm SFR}$}}
\newcommand{\Sgas}{\mbox{$\Sigma_{\rm gas}$}}
\newcommand{\tcc}{\mbox{$\tau_{\rm cc}$}}
\newcommand{\tsf}{\mbox{$\tau_{\rm sf}$}}
\newcommand{\SHI}{\mbox{$\Sigma_{\rm HI}$}}
\newcommand{\kena}{\cite{kennicutt98a}}
\newcommand{\kenb}{\cite{kennicutt98b}}

\thesaurus{3  
          (11.09.1  
           11.19.2  
           11.19.3  
           11.19.4  
           08.06.2  
          )}
\title{Young massive star clusters in nearby spiral galaxies
  \thanks{Based on observations made with the Nordic Optical Telescope,
          operated on the island of La Palma jointly by Denmark, Finland,
          Iceland, Norway, and Sweden, in the Spanish Observatorio del 
          Roque de los Muchachos of the Instituto de Astrofisica de Canarias,
          and with the Danish 1.5-m telescope at ESO, La Silla, Chile.}
}
\subtitle{III. Correlations between cluster populations and host galaxy
 properties}
\author{S.S. Larsen  \inst{1,2}
        \and T. Richtler \inst{3,4}}

\offprints{S.S. Larsen}

\institute{Copenhagen University Astronomical Observatory, 
           Juliane Maries Vej 32, 2100 Copenhagen {\O}, Denmark
        \and UCO/Lick Observatories, Kerr Hall, UC Santa Cruz,
	   CA 95064, USA \\
           email: soeren@ucolick.org
        \and Sternwarte der Universit{\"a}t Bonn,  
           Auf dem H{\"u}gel 71, D-53121 Bonn, Germany
        \and Grupo de Astronom\'{\i}a, Departamento de F\'{\i}sica, \\
	   Casilla 160-C, Universidad de Concepci\'on, Concepci\'on, Chile \\
           email: tom@coma.cfm.udec.cl
	   }

\date{Received ...; accepted ...}

\maketitle
\markboth{Young massive star clusters..}{}

\begin{abstract}
  We present an analysis of correlations between integrated properties
of galaxies and their populations of young massive star clusters.
Data for 21 nearby galaxies presented by Larsen \& Richtler (\cite{lr99}) are 
used together with literature data for 10 additional galaxies, spanning
a range in specific $U$-band cluster luminosity \tlu\ from 0 to 15.
We find that \tlu\ correlates with several observable host galaxy parameters, 
in particular the ratio of Far-Infrared (FIR) to \mbox{$B$-band} flux and 
the surface 
brightness. Taking the FIR luminosity as an indicator of the star formation 
rate (SFR), it is found that \tlu\ correlates very well with the SFR per 
unit area. A similar correlation is seen between \tlu\ and the atomic 
hydrogen surface density.  The cluster formation efficiency seems to depend 
on the SFR in a continuous way, rather than being related to any particularly 
violent mode of star formation.
We discuss fundamental features of possible scenarios for cluster formation.
One possibility is that the correlation between \tlu\ and SFR is due to a 
common controlling parameter, most probably the high density of the ISM.
Another scenario conceives a high \tlu\ as resulting from the energy input from
many massive stars in case of a high SFR. 

\end{abstract}

\section{Introduction}

   A puzzling problem is to understand why different galaxies have 
such widely different young cluster populations as is observed.  The star 
clusters in the Milky Way clearly do not constitute a representative 
cluster sample, as is evident already from a superficial comparison with our
nearest extragalactic neighbours, the Magellanic Clouds. It was noted
early on that the Clouds, in particular the \object{LMC}, contain a 
number of very 
massive, young clusters that do not have any counterparts in our own galaxy 
(van den Bergh \cite{vanden91}; Richtler \cite{richtler93}). Many recent 
studies have shown the presence of such ``Young Massive Clusters'' (YMCs) 
also in a number of mergers and starburst galaxies (see e.g. list in
Harris \cite{harris99}) and it is clear that the occurrence of such objects 
is often associated with violent star formation, leading to the formation of 
a large number of YMCs within a few times $10^8$ years or so. This does 
not explain, however, why other galaxies like the Magellanic Clouds are 
able to maintain the formation of YMCs over a much longer time span. YMCs 
with a broad age distribution have also been found in a few other galaxies, 
e.g. the blue compact galaxy \object{ESO 338-IG04} 
({\"O}stlin et al. \cite{ostlin98}), 
and in the Sc spirals \object{M101} and \object{M33} 
(Bresolin et al. \cite{bresolin96}; Christian \& Schommer \cite{cs88}).

\nocite{bica96}
\nocite{carlson98}
\nocite{oconnell94}
\nocite{johnson99}
\nocite{zepf99}
\nocite{schweizer96}
\nocite{gorjian96}
\nocite{miller97}

\begin{table*}
\caption{Basic properties for the galaxies discussed in this paper.
The data for galaxies labeled $^1$ are taken from the literature
(\object{NGC~1275}: Carlson et al. 1998, 
\object{NGC 1569} / \object{NGC 1705}: O'Connell et al. 1994, 
\object{NGC~1741}: Johnson et al. 1999,
\object{NGC~3256}: Zepf et al. 1999, 
\object{NGC~3921}: Schweizer et al.  1996, 
\object{NGC~5253}: Gorjian et al. 1996, 
\object{NGC 7252}: Miller et al. 1997, 
\object{LMC}: Bica et al. 1996), 
while the remaining data
are from Paper1. The column labeled m-M is the distance modulus 
(see Paper1 for references), $N$ is the number of YMCs, $V_m$ is the $V$ 
magnitude of the brightest cluster, $m_B$ is the integrated $B$
magnitude of the galaxy, $A_B$ is the galactic foreground reddening, and $T_N$ 
is the ``specific frequency''. The two last columns, \tlu\ and $T_L(V)$ give
the specific luminosities of the cluster systems in the $U$ and
$V$ bands.
\label{tab:bprop}
}
\begin{tabular}{lrrrrrrrr} \\ \hline
 Name    & m-M  & $N$ & $V_m$ & $m_B$ & $A_B$ & $T_N$ & \tlu\ & $T_L(V)$ \\ \hline
\multicolumn{9}{c}{Paper1 sample}\\
\object{NGC   45} & 28.4 &   2 & $-9.9 $ & 11.32 & 0.06 & 0.28 & 0.24 & 0.11 \\
\object{NGC  247} & 27.0 &   3 & $-10.2$ &  9.67 & 0.07 & 0.33 & 0.30 & 0.14 \\
\object{NGC  300} & 26.7 &   3 & $ -9.9$ &  8.72 & 0.02 & 0.18 & 0.13 & 0.05 \\
\object{NGC  628} & 29.6 &  39 & $-11.3$ &  9.95 & 0.13 & 0.48 & 0.81 & 0.29 \\
\object{NGC 1156} & 29.5 &  22 & $-11.1$ & 12.32 & 0.71 & 1.61 & 1.67 & 1.08 \\
\object{NGC 1313} & 28.2 &  46 & $-12.1$ &  9.20 & 0.04 & 1.12 & 1.47 & 0.80 \\
\object{NGC 1493} & 30.4 &   0 &   -     & 11.78 & 0.00 & 0.00 & 0.00 & 0.00 \\
\object{NGC 2403} & 27.5 &  14 & $ -9.9$ &  8.93 & 0.17 & 0.45 & 0.24 & 0.14 \\
\object{NGC 2835} & 28.9 &  12 & $-10.9$ & 11.01 & 0.44 & 0.57 & 0.55 & 0.30 \\
\object{NGC 2997} & 29.9 &  34 & $-12.9$ & 10.06 & 0.54 & 0.25 & 1.45 & 0.99 \\
\object{NGC 3184} & 29.5 &  13 & $-10.6$ & 10.36 & 0.00 & 0.28 & 0.23 & 0.10 \\
\object{NGC 3621} & 29.1 &  51 & $-11.9$ & 10.18 & 0.42 & 0.93 & 1.33 & 0.65 \\
\object{NGC 4395} & 28.1 &   2 & $-9.1 $ & 10.64 & 0.01 & 0.21 & 0.07 & 0.05 \\
\object{NGC 5204} & 28.4 &   7 & $-9.6 $ & 11.73 & 0.00 & 1.49 & 0.39 & 0.38 \\
\object{NGC 5236} & 27.9 & 153 & $-11.7$ &  8.20 & 0.15 & 1.77 & 2.33 & 0.90 \\
\object{NGC 5585} & 29.2 &   7 & $-10.8$ & 11.20 & 0.00 & 0.44 & 0.50 & 0.31 \\
\object{NGC 6744} & 28.5 &  18 & $-11.0$ &  9.14 & 0.15 & 0.28 & 0.51 & 0.14 \\
\object{NGC 6946} & 28.7 & 107 & $-13.0$ &  9.61 & 1.73 & 0.56 & 1.44 & 0.58 \\
\object{NGC 7424} & 30.5 &   9 & $-11.4$ & 10.96 & 0.00 & 0.14 & 0.38 & 0.19 \\
\object{NGC 7741} & 30.8 &   0 &   -     & 11.84 & 0.15 & 0.00 & 0.00 & 0.00 \\
\object{NGC 7793} & 27.6 &  20 & $-10.4$ &  9.63 & 0.02 & 1.21 & 1.15 & 0.51 \\
\multicolumn{9}{c}{Starbursts / mergers}\\
\object{NGC 1275}$^1$ & 34.2 &  -  & $-14  $ & 12.64 & 0.75 &  -   & 2.63 & 1.04 \\
\object{NGC 1569}$^1$ & 27.0 &  -  & $-13.9$ & 11.86 & 2.18 &  -   & 11.3 & 5.60 \\
\object{NGC 1705}$^1$ & 28.5 &  -  & $-13.7$ & 12.77 & 0.19 &  -   & 13.9 & 10.1 \\
\object{NGC 1741}$^1$ & 33.5 &  -  & $-15  $ & 13.30 & 0.25 &  -   & $\sim10$ & $\sim5$ \\
\object{NGC 3256}$^1$ & 32.8 &  -  & $-15  $ & 12.15 & 0.59 &  -   & $\sim15$ & $\sim15$ \\
\object{NGC 3921}$^1$ & 36.0 &  -  & $-14  $ & 13.06 & 0.16 &  -   & 0.24 & 0.11 \\
\object{NGC 5253}$^1$ & 28.0 &  -  & $-11.1$ & 10.87 & 0.20 &  -   & 1.41 & 0.51 \\
\object{NGC 7252}$^1$ & 34.9 &  -  & $-17.0$ & 12.06 & 0.05 &  -   & 2.43 & 1.10 \\ 
\multicolumn{9}{c}{Other galaxies}\\
\object{IC  1613}$^1$ & 24.3 &  -  &  -    &  9.88 & 0.02 &  0   &  0   &  0 \\
\object{LMC}$^1$  & 18.5 &   8 &  $-9.4$ &  0.91 & 0.27 & 0.57 & 0.12 & 0.11 \\ 
\hline
\end{tabular}
\end{table*}

\nocite{rice88}
\nocite{soifer89}

\begin{table*}
\caption{
  Integrated properties for the galaxies, mostly taken from the RC3
catalogue. T is the revised Hubble type, coded as in RC3. $m_{25}$ is the 
average \mbox{$B$-band} surface brightness 
within an ellipse corresponding to 25 mag / square arc second, and
$\log D_0$ is the face-on diameter corrected for galactic extinction.
$m_{21}$ is a magnitude derived from the 21-cm flux.
\mfir\ is a Far-Infrared magnitude based on the IRAS $60\mu$ and 
$100\mu$ fluxes. The IRAS 60 and 100$\mu$ fluxes are in units of Jy. 
\Ssfr\ (given as $10^3 \times$ \msun\ yr$^{-1}$ kpc$^{-2}$) and \SHI\ 
(in units of \msun\ pc$^{-2}$) are derived from \mfir\ and $m_{21}$ as 
described in Sect.~\ref{sec:corr}.
$^1$: FIR data from Rice et al. (1988) 
$^2$: FIR data from Soifer et al. (1989) 
$^3$: FIR data from IRAS Faint Source Catalog (1990) 
$^4$: B-V and U-B from RC3 
$^5$: U-B measured by us.
See the text for further explanation.
\label{tab:intdata}
}
\begin{tabular}{lrrrrrrrrrp{6mm}r} \hline
  Name            & T   & U-B   & B-V   &$m_{25}$ & $m_{21}$ & \mfir\ & $f(60\mu)$ & $f(100\mu)$ &$\log D_0$ & \Ssfr\ \mbox{$\times 10^3$} & \SHI\ \\ 
\hline
\multicolumn{12}{c}{Paper1 sample}\\
\object{NGC    45}$^{1,4}$ & 8.0 & $-0.05$ &  0.71 & 15.39   & 11.43 &  12.34 &  1.62 & 4.99 & 1.93 & 0.23 & 12.1 \\ 
\object{NGC   247}$^{1,5}$ & 7.0 & $-0.10$ &  0.56 & 14.95   & 10.27 &  10.55 &  7.93 & 27.3 & 2.34 & 0.18 & 5.3 \\ 
\object{NGC   300}$^{1,4}$ & 7.0 &  0.11 &  0.59 & 14.91   &  9.15 &   9.43 &  23.1 & 74.4 & 2.35 & 0.49 & 14.3 \\ 
\object{NGC   628}$^{1,5}$ & 5.0 &  0.00 &  0.56 & 14.79   & 10.77 &   9.56 &  20.9 & 65.6 & 2.03 & 1.88 & 14.0 \\ 
\object{NGC  1156}$^{3,4}$ & 10.0 &$ -0.19$ & 0.58 & 14.43   & 12.72 &  11.28 &  5.71  & 9.20 & 1.58 & 3.07 & 18.4 \\ 
\object{NGC  1313}$^{1,4}$ & 7.0 & $-0.24$ &  0.49 & 13.52   & 10.54 &   9.08 &  36.0 & 92.0 & 1.96 & 4.04 & 23.9 \\ 
\object{NGC  1493}$^{3,5}$ & 6.0 & $-0.06$ &  0.52 & 14.27   & 13.38 &  11.89 &  2.33 & 8.19  & 1.54 & 2.10 & 12.1 \\ 
\object{NGC  2403}$^{1,5}$ & 6.0 & $-0.10$ &  0.47 & 14.88   &  9.58 &   8.63 &  51.6 & 148  & 2.36 & 0.97 &  9.2 \\ 
\object{NGC  2835}$^{3,4}$ & 5.0 & $-0.12$ &  0.49 & 14.51   & 11.98 &  11.44 &  3.25  & 16.0 & 1.86 & 0.73 & 10.0 \\ 
\object{NGC  2997}$^{1,5}$ & 5.0 &  0.10 &  0.00 & 14.33   & 11.50 &   9.18 &  32.3 & 85.1 & 2.00 & 3.07 &  8.2 \\ 
\object{NGC  3184}$^{2,4}$ & 6.0 & $-0.03$ &  0.58 & 14.49   & 12.18 &  10.46 &  8.92 & 29.0 & 1.87 & 1.72 &  8.0 \\ 
\object{NGC  3621}$^{1,4}$ & 7.0 & $-0.08$ &  0.62 & 14.90   & 10.20 &   9.19 &  29.6 & 90.1 & 2.13 & 1.67 & 14.9 \\ 
\object{NGC  4395}$^{1,5}$ & 9.0 &  0.10 &  0.46 & 15.86   & 11.11 &  11.31 &  4.21 & 12.9 & 2.12 & 0.25 &  6.8 \\ 
\object{NGC  5204}$^{3,4}$ & 9.0 & $-0.33$ &  0.41 & 14.55   & 12.35 &  12.10 &  2.32 & 5.35  & 1.70 & 0.83 & 14.9 \\ 
\object{NGC  5236}$^{1,4}$ & 5.0 &  0.03 &  0.66 & 13.48   &  9.60 &   6.95 &  266  & 639  & 2.12 & 13.8 & 27.2 \\ 
\object{NGC  5585}$^{3,4}$ & 7.0 & $-0.22$ &  0.46 & 14.35   & 12.10 &  12.82 &  0.99 & 3.65  & 1.76 & 0.33 & 14.3 \\ 
\object{NGC  6744}$^{1,5}$ & 4.0 &  0.13 &  0.75 & 15.00   &  9.55 &   9.36 &  22.2 & 85.8 & 2.31 & 0.62 & 11.9 \\ 
\object{NGC  6946}$^{1,5}$ & 6.0 &  0.20 &  0.80 & 14.58   & 10.09 &   7.64 &  137  & 344  & 2.22 & 4.60 & 10.9 \\ 
\object{NGC  7424}$^{1,3}$ & 6.0 & $-0.15$ &  0.48 & 15.52   & 11.27 &  12.36 &  1.22  & 7.83  & 1.98 & 0.18 & 11.1 \\ 
\object{NGC  7741}$^{3,4}$ & 6.0 & $-0.14$ &  0.53 & 14.45   & 13.15 &  12.00 &  2.27  & 6.98 & 1.65 & 1.14 &  9.0 \\ 
\object{NGC  7793}$^{3,4}$ & 7.0 & $-0.09$ &  0.54 & 13.91   & 11.21 &   9.68 &  19.6 & 56.3 & 1.98 & 2.12 & 11.7 \\ 
\multicolumn{12}{c}{Starbursts / mergers}\\
\object{NGC  1275}$^{3,4}$ & -   & 0.07  &  0.76 &  -      &  -    &  11.24 &  7.15 & 6.98 & 1.41 & 6.96 & - \\
\object{NGC  1569}$^{3,4}$ & 10.0 & $-0.14$ &  0.83 & 13.71  & 12.43 &   9.16 &  45.4 & 47.3 & 1.76 & 9.43 & 10.5 \\ 
\object{NGC  1705}$^{3,4}$ & $-3.0$ & $-0.45$ &  0.38 & 13.70  &   -   &  13.16 &  0.87  & 1.82  & 1.28 & 2.16 & - \\ 
\object{NGC  1741}$^{3,-}$ & 10  &  -    &   -   &  -      & 14.03 &  11.73 &  3.92  & 5.84  & 1.18 & 12.8 & 34.8 \\
\object{NGC  3256}$^{3,4}$ & -   & $-0.08$ &  0.64 &  -      &  -    &   8.34 &  88.3 & 115 & 1.63 & 36.5 & - \\
\object{NGC  3921}$^{3,4}$ & 2  &  0.25 &  0.68 & 14.04   & 15.51  &  13.29 &  0.83  & 0.0 & 1.32 & 1.59 & 4.7 \\ 
\object{NGC  5253}$^{3,4}$ & 10  & $-0.24$ &  0.43 & 13.18   & 13.00 &   9.64 &  30.5 & 29.4 & 1.72 & 7.29 & 7.5 \\
\object{NGC  7252}$^{3,4}$ & -    &  0.20 &  0.66 & 13.83  &  -    &  11.60 &  3.98 & 7.02  & 1.28 & 9.09 & - \\ 
\multicolumn{12}{c}{Other galaxies}\\
\object{IC 1613}$^{3,4}$   &10.0 &   -   &  0.67 & 15.68   & 10.73 &  12.58 &  0.98   & 2.67      & 2.22 & 0.05 & 6.1 \\
\object{LMC}$^{2,4}$       & 9.0 &  0.00 &  0.51 & 14.64   &  2.75 &   0.74 &  82900 & 185000 & 3.84 & 1.51 &  5.4 \\ 
\hline
\end{tabular}
\end{table*}

In Larsen \& Richtler (\cite{lr99}, hereafter Paper1) we carried out a 
systematic search for YMCs in 21 nearby non-interacting, mildly inclined 
galaxies, and identified rich populations of YMCs in about a quarter of the 
galaxies in the sample. Within the range of Hubble types surveyed (Sbc -- Irr),
no correlation was found between the morphological type of the galaxies and 
their contents of YMCs.  In the present paper we show that the richness of 
the cluster systems is indeed well correlated with certain other properties of 
the host galaxies, indicative of a dependence on the star formation rate.  We 
extend our sample relative to Paper1 by also including literature data for a 
variety of different star-forming galaxies, and show that the correlations 
inferred from our sample are further strengthened when the additional data 
are included.  Hence, it seems that starburst galaxies with their very 
rich populations of YMCs represent only an extreme manifestation of the 
cluster formation process, while the conditions that allow YMCs to be formed 
can be present also in normal galaxies. 

\section{Basic definitions}
\label{sec:defs}

  The data reduction procedure and identification of YMCs have been
discussed elsewhere (Paper1; Larsen \cite{larsen99}) and we shall not 
repeat the 
details here. We just mention that the clusters were identified using 
broad-band photometry, applying a colour criterion of $B-V < 0.45$ (mainly 
in order to exclude foreground stars) and an absolute visual magnitude limit 
of $M_V = -8.5$ for objects with $U-B >= -0.4$ and $M_V = -9.5$ for 
$U-B < -0.4$. The $B-V$ colour cut-off corresponds to an age of about 500 Myr
(Girardi et al. \cite{girardi95}) and the lower mass limit is of the
order of $3 \times 10^4$ \msun , assuming a Salpeter IMF extending down
to 0.1 \msun\ (Bruzual \& Charlot \cite{bc93}).
``Fuzzy'' objects and HII regions were excluded by a combination of
visual inspection and H$\alpha$ photometry (see Larsen \cite{larsen99}
for details). Hence, we define an object that satisfies these criteria to 
be a Young Massive Cluster.

  Following the definition of the ``specific frequency'' $S_N$ for old
globular cluster systems (Harris \& van den Bergh \cite{hv81}), we defined an
equivalent quantity for young clusters in Paper1:

\begin{equation}
  T_N = N \times 10^{0.4 \times (M_B + 15)}
  \label{eq:tn}
\end{equation}

  Here $N$ is the total number of YMCs in a galaxy, and $M_B$ is the
absolute $B$ magnitude of the galaxy. $T_N$ is then a measure of 
the number of clusters, normalised to the luminosity of the host galaxy.
There are, however, several problems in defining a ``specific frequency''
for young clusters. Since old globular cluster systems have a log-normal like
luminosity function (LF), the total number of old clusters belonging to a 
given galaxy is a well-defined quantity, and can be estimated with good
accuracy even if the least luminous clusters cannot be observed 
directly. Young clusters, on the other hand, usually exhibit a
power-law luminosity function of the form
\begin{equation}
  N(L)dL \propto L^{-\alpha} dL
  \label{eq:ymclf}
\end{equation}
with an increasing number of clusters at
fainter magnitudes. Hence, $T_N$ depends sensitively on the definition
one adopts for a YMC, and it is difficult to compare literature data
unless the exact selection parameters are known. Moreover, incompleteness
effects and errors in the distance modulus always affect the number
of clusters in the faintest magnitude bins most severely, and this
leads to large uncertainties in $T_N$.

  Another possibility is to consider the total {\it luminosity} of the
cluster system compared to that of the host galaxy. This approach has the
advantage of being independent of the distance modulus and interstellar
absorption. Following Harris (\cite{harris91}), we define the 
{\it specific luminosity}
\begin{equation}
  T_L = 100 \cdot \frac{L_{\rm Clusters}}{L_{\rm Galaxy}} 
  \label{eq:tu}
\end{equation}
  where $L_{\rm Clusters}$ and $L_{\rm Galaxy}$ are the total luminosities
of the cluster system and of the host galaxy, respectively.  It makes no 
difference if the absolute or apparent luminosities are used in 
Eq.~(\ref{eq:tu}), and corrections for reddening only play a role through the 
selection criteria for identification of YMCs. 

  As long as the exponent $\alpha$ in the LF (Eq. (\ref{eq:ymclf})) is less 
than 2, most of the light originates from the {\it bright} end of the LF. 
A typical value is $\alpha \approx 1.7$ (Elmegreen \& Efremov \cite{ee97}; 
Harris \& Pudritz \cite{hp94}), although slopes of $\alpha \sim 2$ have also 
been reported (e.g. for NGC~3921, Schweizer et al. \cite{schweizer96}). In
any case, $T_L$ is much less sensitive to incompleteness effects at the
lower end of the LF than the specific frequency.

  We remark that the brighter end of the LF of old globular cluster systems 
is also well described by a power-law distribution with an exponent
similar to that observed for the young cluster populations. This has 
stimulated attempts to create a universal theoretical description of the 
formation of old globular clusters in the halo of the Milky Way and elsewhere
as well as the present-day formation of young star clusters (Elmegreen
\& Efremov \cite{ee97}; McLaughlin \& Pudritz \cite{mp96}).

\section{The data}

  The basic data related to the cluster systems considered in this paper
are given in Table~\ref{tab:bprop}. The number of YMCs $N$ and corresponding
specific frequencies $T_N$ are taken from Paper1, and in addition 
we now also list the absolute \mbox{$V$-band} magnitude of the brightest 
cluster in each galaxy $V_m$ and the $U$- and $V$-band specific luminosities
\tlu\ and \tlv .  The $T_N$ values in Tables~\ref{tab:bprop} have not been 
corrected for completeness effects, which can be quite significant in 
particular for the more distant galaxies like \object{NGC~2997} (Larsen 
\cite{larsen99}). However, we are
not going to refer much to $T_N$ in this paper for the reasons given in
Sect.~\ref{sec:defs} but will instead use specific luminosities.
We remark that the often very luminous clusters found near the centres of 
certain ``hot spot'' galaxies (e.g.  \object{NGC~2997}, Maoz et al. 
\cite{maoz96} and \object{NGC~5236}, Heap et al. \cite{heap93}) have not 
been considered in this study, but only clusters in the disks.

  In addition to the Paper1 sample, we also include literature data for a 
number of (mostly) starburst and merger galaxies (see references 
in the caption to Table~\ref{tab:bprop}).  Since the clusters in these 
galaxies were not identified according to a homogeneous set of criteria we 
do not list $T_N$ values, except for the LMC 
where the published photometry reaches below $M_V = -8.5$. 
The photometry published for clusters in the 
remaining galaxies does not go as deep as ours but as we have argued above, 
the total integrated magnitude of a cluster system is normally dominated by 
the brighter clusters, so we have calculated \tlu\ and \tlv\ values for all 
galaxies based on the available data. Not all studies list $UBV$ colours,
but these have been estimated from the published cluster ages and the
Girardi et al. (\cite{girardi95}) ``S''-sequence.  

  Table~\ref{tab:intdata} lists integrated data for the galaxies, mostly
taken from the RC3 catalogue, with the exception of the $U-B$ colour which
has in a few cases been derived from our own CCD data.
$T$ is the revised Hubble type, $m_{25}$ is the \mbox{$B$-band} surface 
brightness, $m_{21}$ is a magnitude based on the 21 cm flux
(see RC3 for details) and \mfir\ is a FIR magnitude based on the IRAS 
fluxes at $60\mu$ and $100\mu$. $\log D_0$ is the logarithm of the face-on 
diameter of the galaxy, and the last two columns in Table~\ref{tab:intdata} 
list the area-normalised star formation rate \Ssfr\ and the HI surface 
density \SHI\ derived from \mfir\ and $m_{21}$ (see 
Sect.~\ref{subsec:p1sample}).  The RC3 as well as the IRAS data were 
retrieved through the NASA/IPAC Extragalactic Database.

\section{Correlations between host galaxy parameters and cluster systems}
\label{sec:corr}

  In this section we discuss correlations between various host galaxy 
properties and the specific $U$-band luminosity \tlu . We use \tlu\ because 
the $U$-band most cleanly samples the {\it young} stellar populations in a 
galaxy, and therefore provides the purest measure of {\it current} cluster 
formation activity.

\subsection{The Paper1 sample}
\label{subsec:p1sample}

\begin{figure}
\begin{minipage}{88mm}
\epsfxsize=88mm
\epsfbox[87 368 555 640]{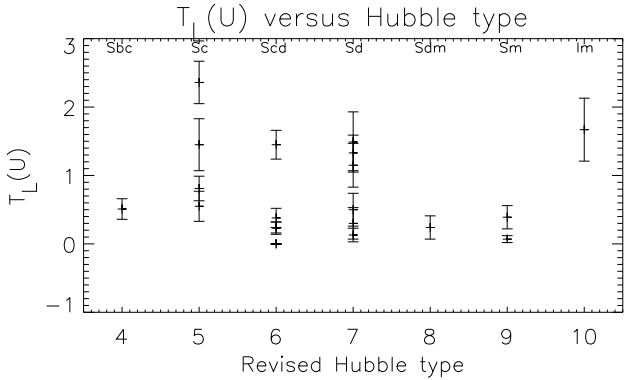}
\end{minipage}
\caption{\label{fig:t_tlu} The specific $U$-band luminosity \tlu\ as a 
function of revised Hubble type.
}
\end{figure}

\begin{figure*}
\begin{minipage}{88mm}
\epsfxsize=88mm
\epsfbox[87 368 555 640]{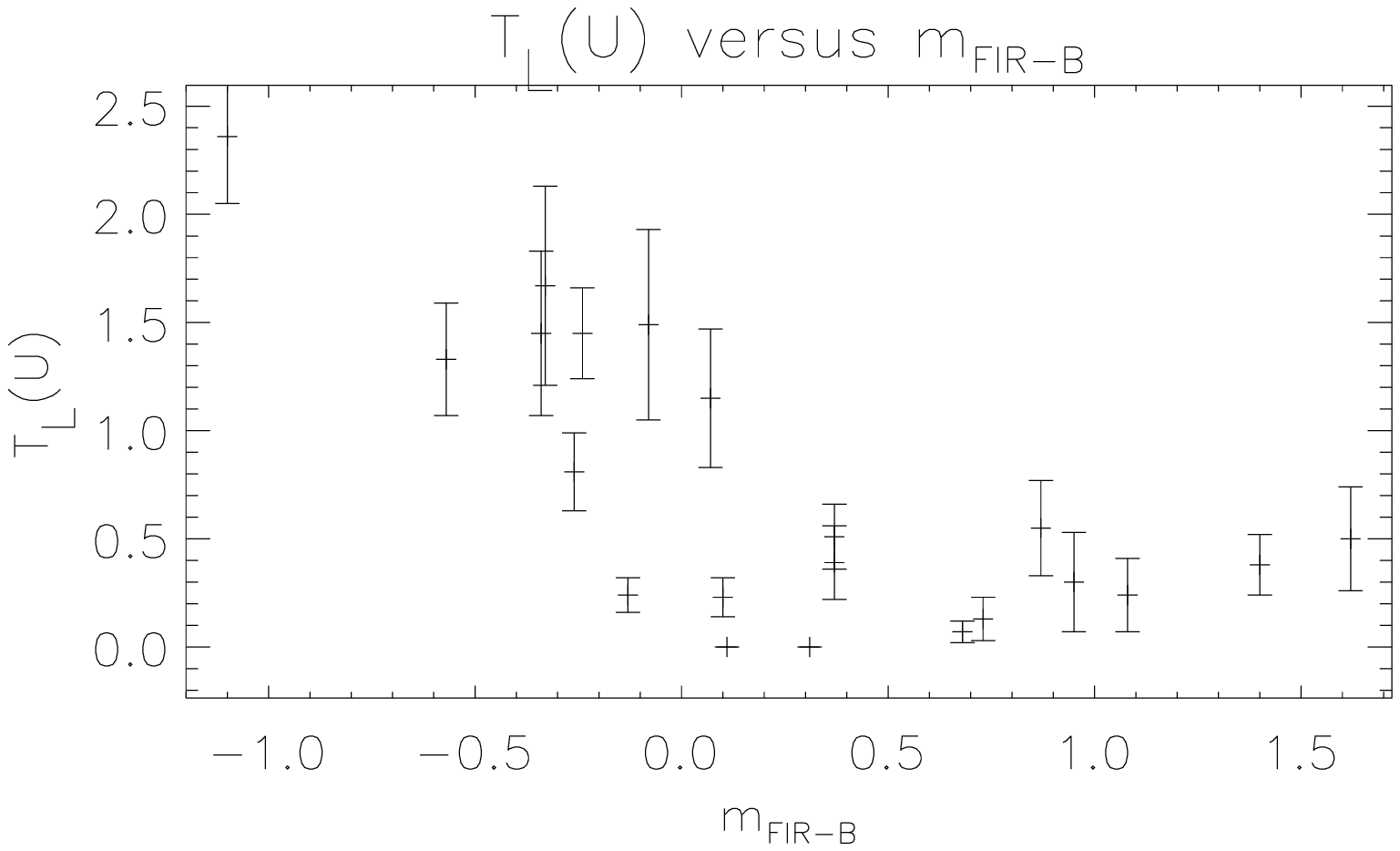}
\end{minipage}
\begin{minipage}{88mm}
\epsfxsize=88mm
\epsfbox[87 368 555 640]{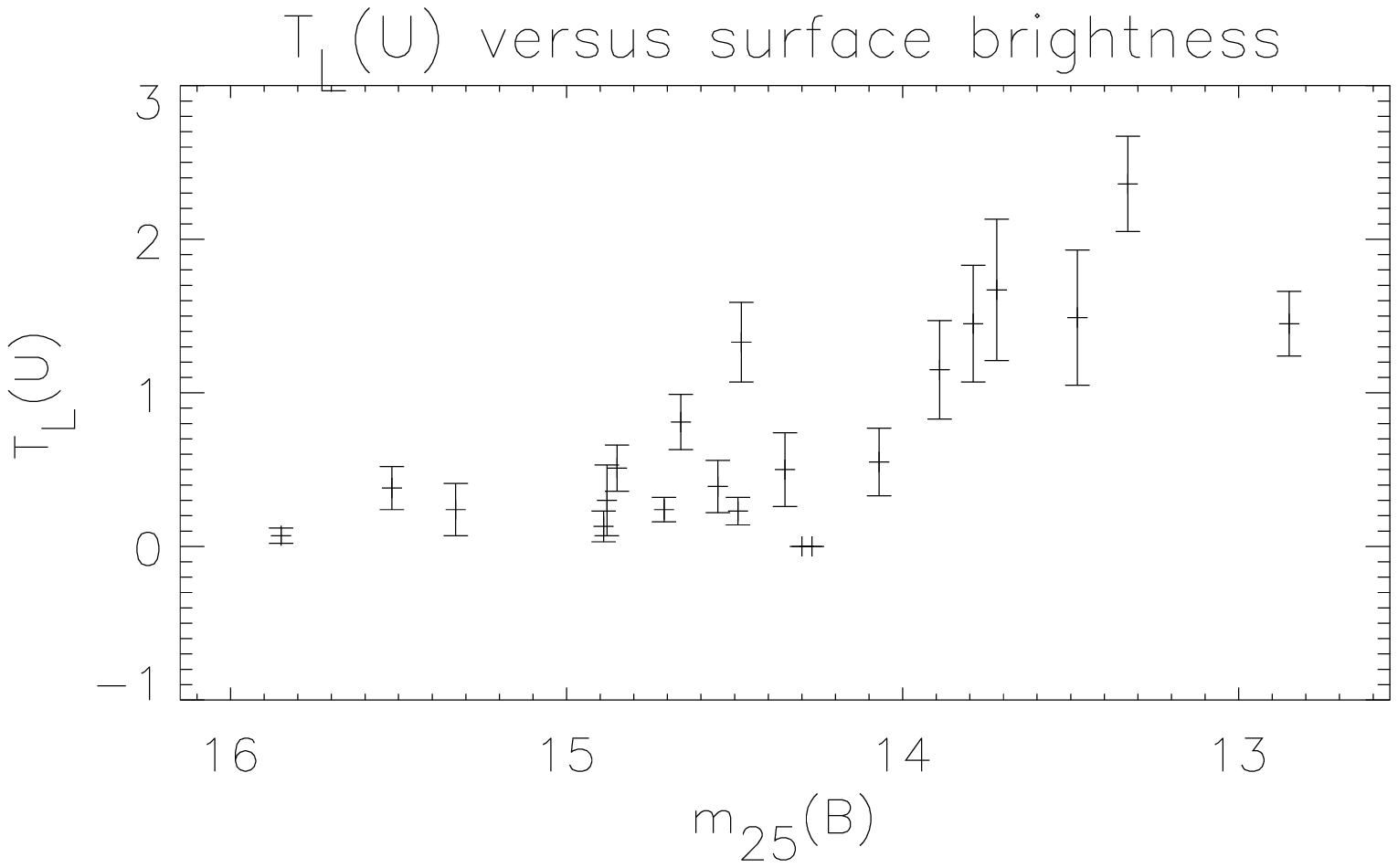}
\end{minipage}
\\
\begin{minipage}{88mm}
\epsfxsize=88mm
\epsfbox[87 368 555 640]{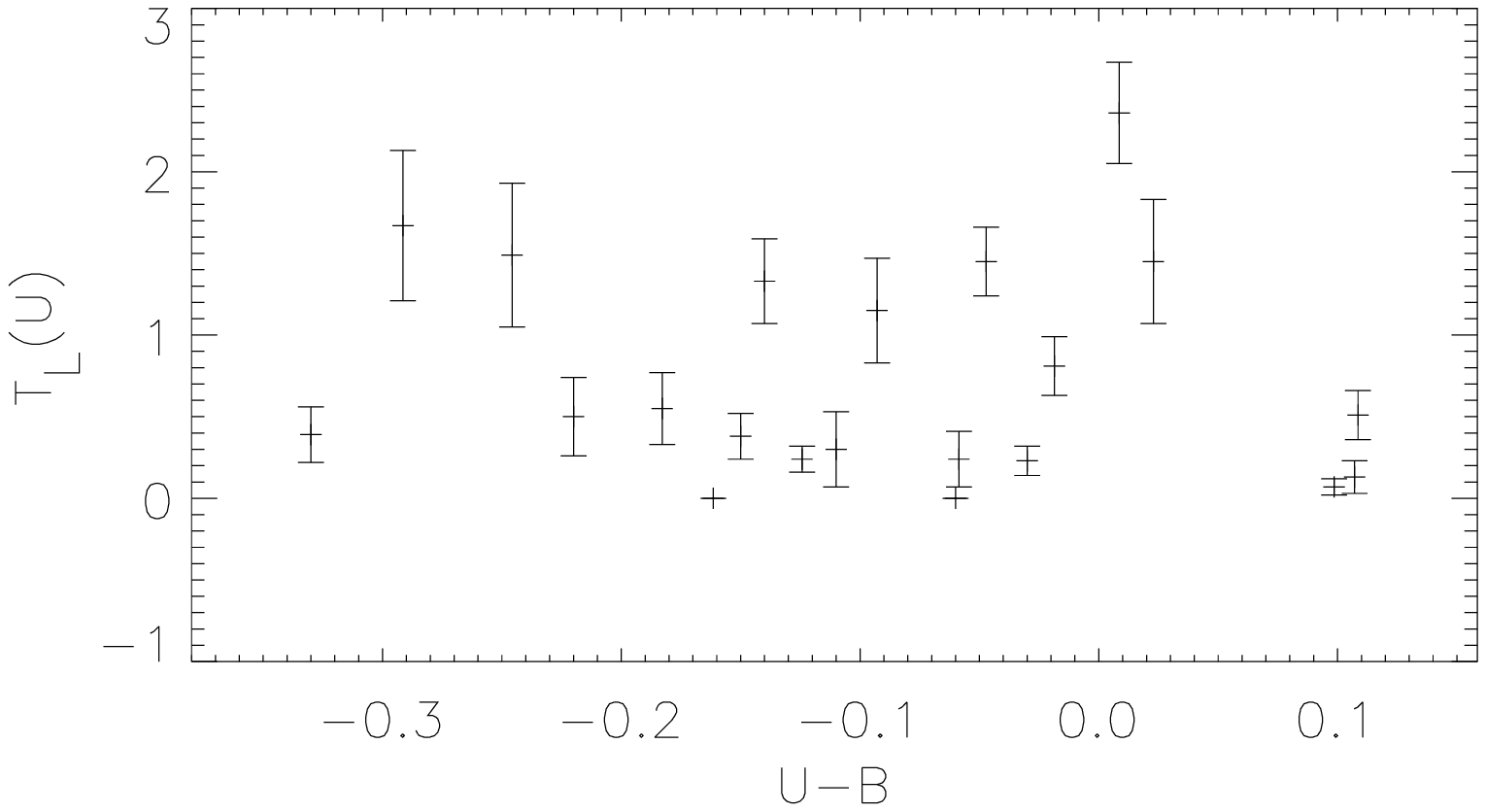}
\end{minipage}
\begin{minipage}{88mm}
\epsfxsize=88mm
\epsfbox[87 368 555 640]{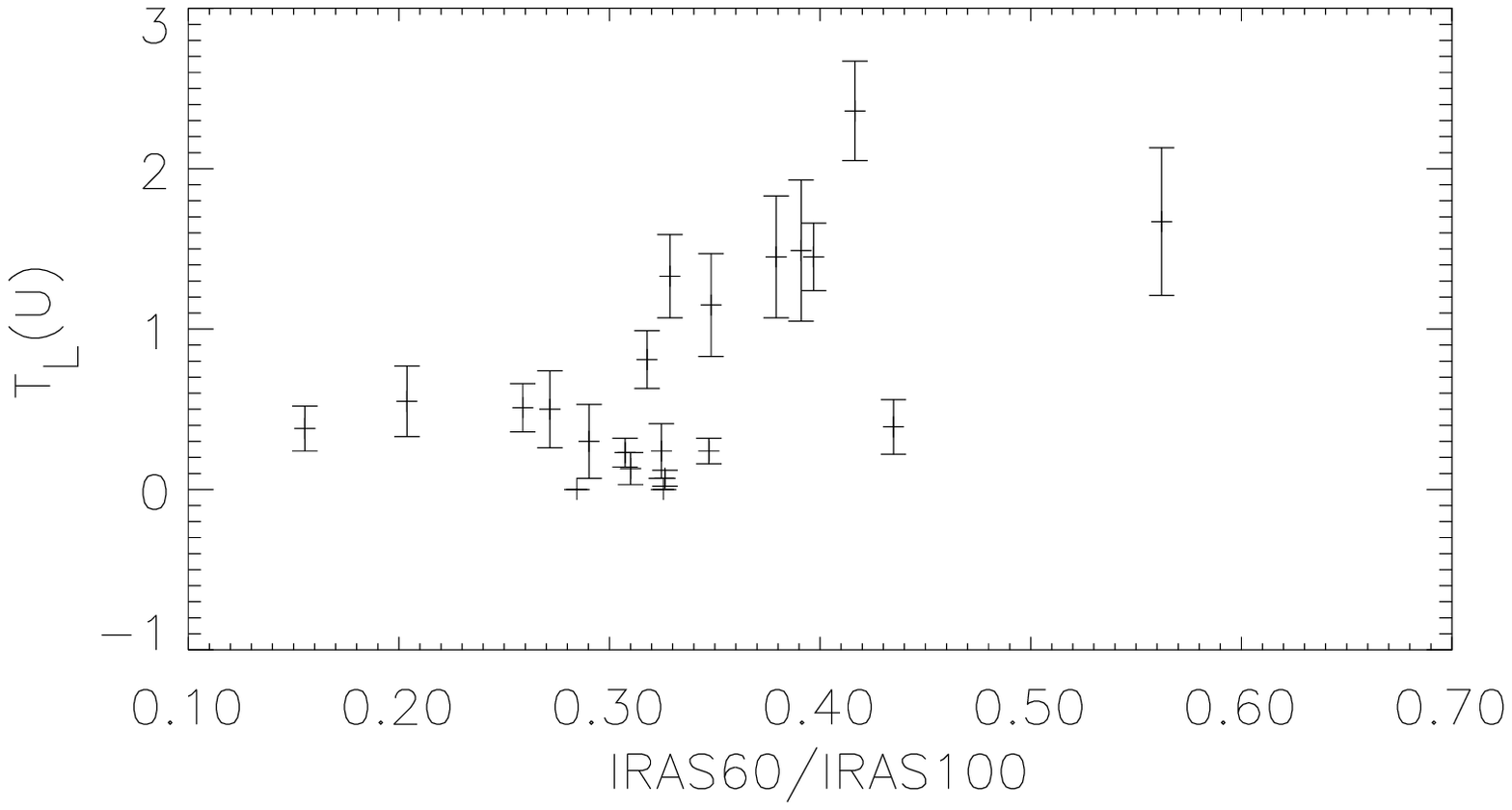}
\end{minipage}
\caption{
  The correlation between various integrated galaxy properties and \tlu .
  \label{fig:tu1}
}
\end{figure*}

  First, we consider only the galaxies studied in Paper1. In Paper1 we
showed that there is no evident correlation between $T_N$ and the Hubble 
type of the host galaxy.  In Fig.~\ref{fig:t_tlu} we show \tlu\ instead 
of $T_N$ as a function of the Hubble type, but this does not change the 
conclusion - there is no clear trend in \tlu\ as a function of Hubble 
type either.  The earliest type represented in our sample is Sbc (type 
4.0 in the RC3 terminology), and the latest is Im (type 10.0 in RC3).  
Independently of morphological type, we find a range from galaxies with 
practically no YMCs to very rich cluster systems in our sample, so even 
if YMCs might be systematically absent in galaxies of even earlier types, 
the presence of YMCs cannot be entirely related to morphology. Furthermore, 
some of the galaxies with high \tlu\ values are grand-design spirals 
(\object{NGC~5236}, \object{NGC~2997}), other grand-design spirals 
are relatively cluster-poor (e.g. \object{NGC~3184}, \object{NGC~7424}),
while the flocculent galaxy \object{NGC~7793} also has a high \tlu\ value,
so the presence of a spiral density wave is apparently not a discriminating 
factor either. No galaxies of types Sa and Sb were included in our sample,
primarily because of a general lack of sufficiently nearby galaxies of these 
types (see Paper1 for a more detailed discussion of the selection criteria).

  We therefore continue to look for other host galaxy parameters that
could correlate with \tlu .  Even for the
relatively nearby galaxies in our sample, it is not an easy task to find
homogeneous sets of observations of integrated properties that allow a
comparison of all galaxies, mainly because the most complete data exist 
for the northern hemisphere while many of our galaxies are in the southern 
sky. For example, existing CO surveys have included only few of our 
galaxies (Elfhag et al. \cite{elfhag96}; Young et al. \cite{young95}), 
We are therefore largely limited to discussing optical data, HI data and 
Far-Infrared data from the IRAS survey.  

  In order to reach independence of distance and absolute galaxy luminosity,
we normalise the FIR flux to the $B$-band magnitude of a galaxy 
by using the ``FIR -- B'' index \mfirmb\ = $m(\mbox{FIR}) - m(B)$.

  Fig.~\ref{fig:tu1} shows \tlu\ as a function of various integrated host 
galaxy parameters: The \mfirmb\ index, the $B$-band surface brightness, 
the integrated $U-B$ colour and the IRAS \ic\ flux ratio.
The $U-B$ and the $B$ band data have been corrected for Galactic foreground 
extinction (as given in Table~\ref{tab:bprop}), but not for internal 
absorption in the galaxies. The latter correction would move the points 
around slightly, but neither reduce the scatter significantly nor change the 
conclusions. We have therefore avoided to apply this anyway quite uncertain 
correction.

  From Fig.~\ref{fig:tu1} we first note the striking correlations between 
\tlu\ and \mfirmb\ and the surface brightness $m_{25}$. This is of great 
interest because both these parameters can be taken as indicators for the 
star formation rate in the host galaxy (Kennicutt \kenb).
We stress that, because we are operating with specific luminosities this is 
not just a sampling effect - Fig.~\ref{fig:tu1} shows that the 
{\it relative} amount of light that originates from clusters, relative to 
the general field population, increases as a function of $m_{25}$ and 
\mfirmb . A similar result regarding surface brightness and the fraction
of UV light in clusters was also noted by Meurer et al. (\cite{meurer95})
for a number of starburst galaxies, but over a smaller range in
surface brightness. 

  No correlation between \tlu\ and the $U-B$ colour is seen, but this may 
not be quite as surprising since the $U-B$ colour index has a less clear 
physical interpretation and is, in any case, severely affected by 
absorption effects.

  There is also some correlation between \tlu\ and the \ic\ flux ratio,
which measures the dust temperature and can therefore be taken as a measure 
of the intensity of the radiation field in a galaxy 
(Soifer et al. \cite{soifer89}).  The radiation field might play a role for
the formation of bound clusters by keeping proto-cluster clouds in thermal
equilibrium and delaying thermal instabilities, thereby preventing 
star formation from setting in too early and disrupting the clouds
(Murray \& Lin \cite{ml92}). 
However, a high \ic\ ratio follows naturally from a high global FIR 
luminosity (Soifer et al. \cite{soifer89}), and the correlation between 
\tlu\ and \ic\ does not by itself provide any evidence that the radiation 
field is a dominating factor in determining whether YMCs can form in a galaxy.
Thermal instabilities might be prevented in other ways, particularly
by magnetic pressure support (see e.g. Mouschovias \cite{mous91}).

\begin{figure}
\begin{minipage}{88mm}
\epsfxsize=88mm
\epsfbox[87 368 555 640]{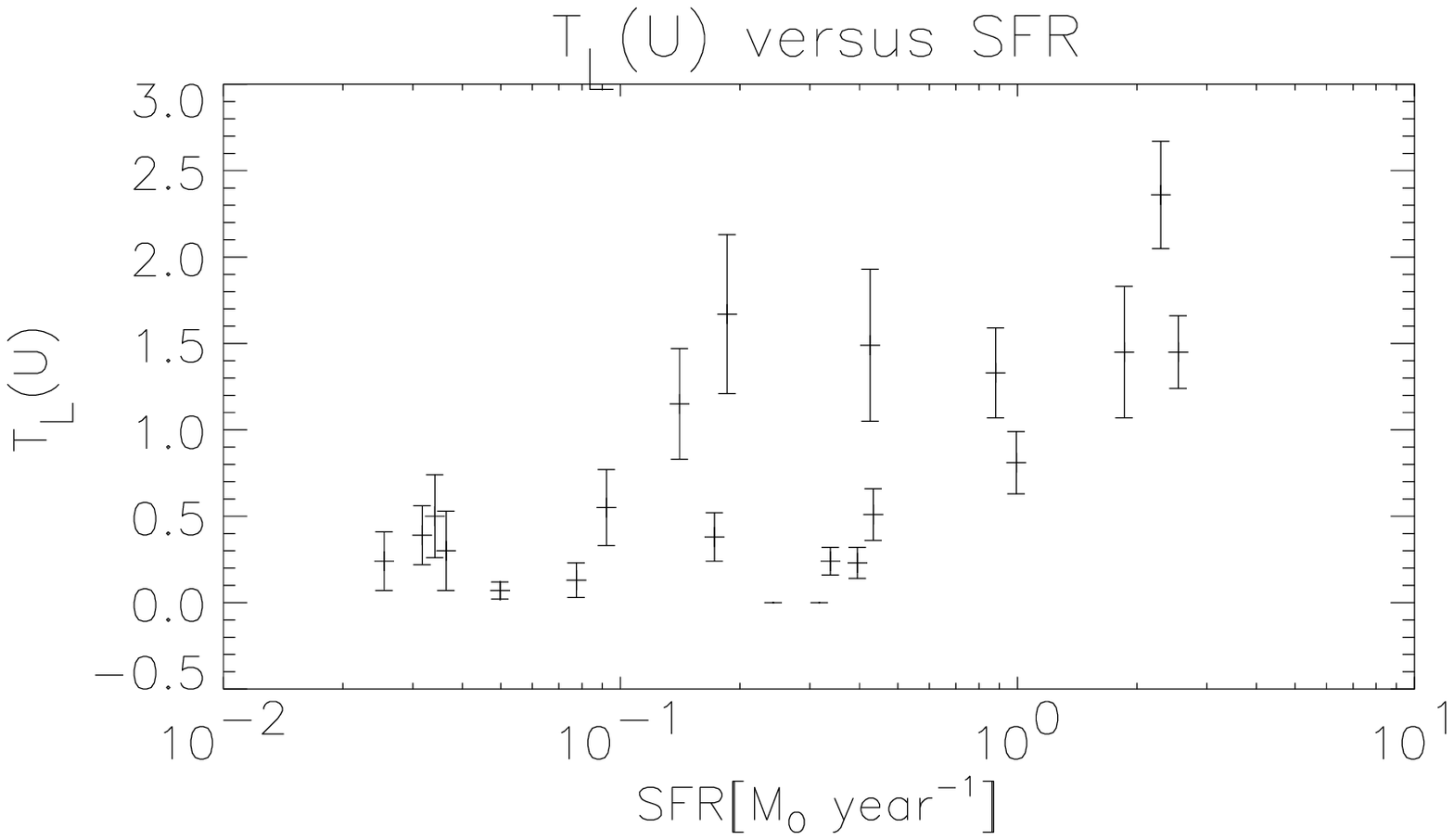}
\end{minipage}
\begin{minipage}{88mm}
\epsfxsize=88mm
\epsfbox[87 368 555 640]{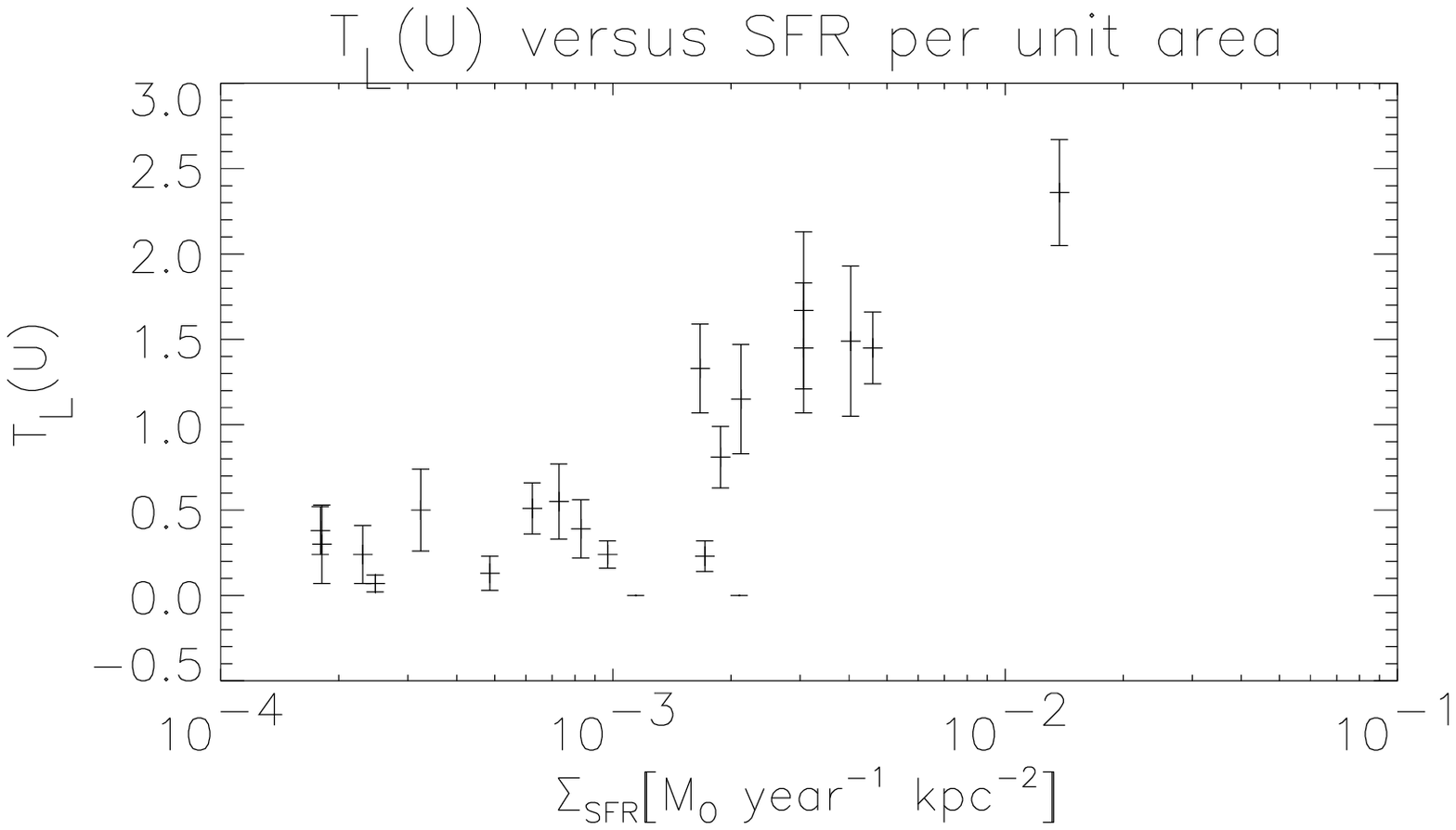}
\end{minipage}
\caption{\label{fig:sfr_p1}
  \tlu\ vs. Star Formation Rate as derived from the FIR luminosities for 
galaxies in the Paper1 sample. The upper panel shows \tlu\ as a function of
the global SFR, while the lower panel shows \tlu\ vs. \Ssfr ,
the SFR per unit area.
}
\end{figure}

\begin{figure}
\epsfxsize=88mm
\epsfbox[87 368 555 640]{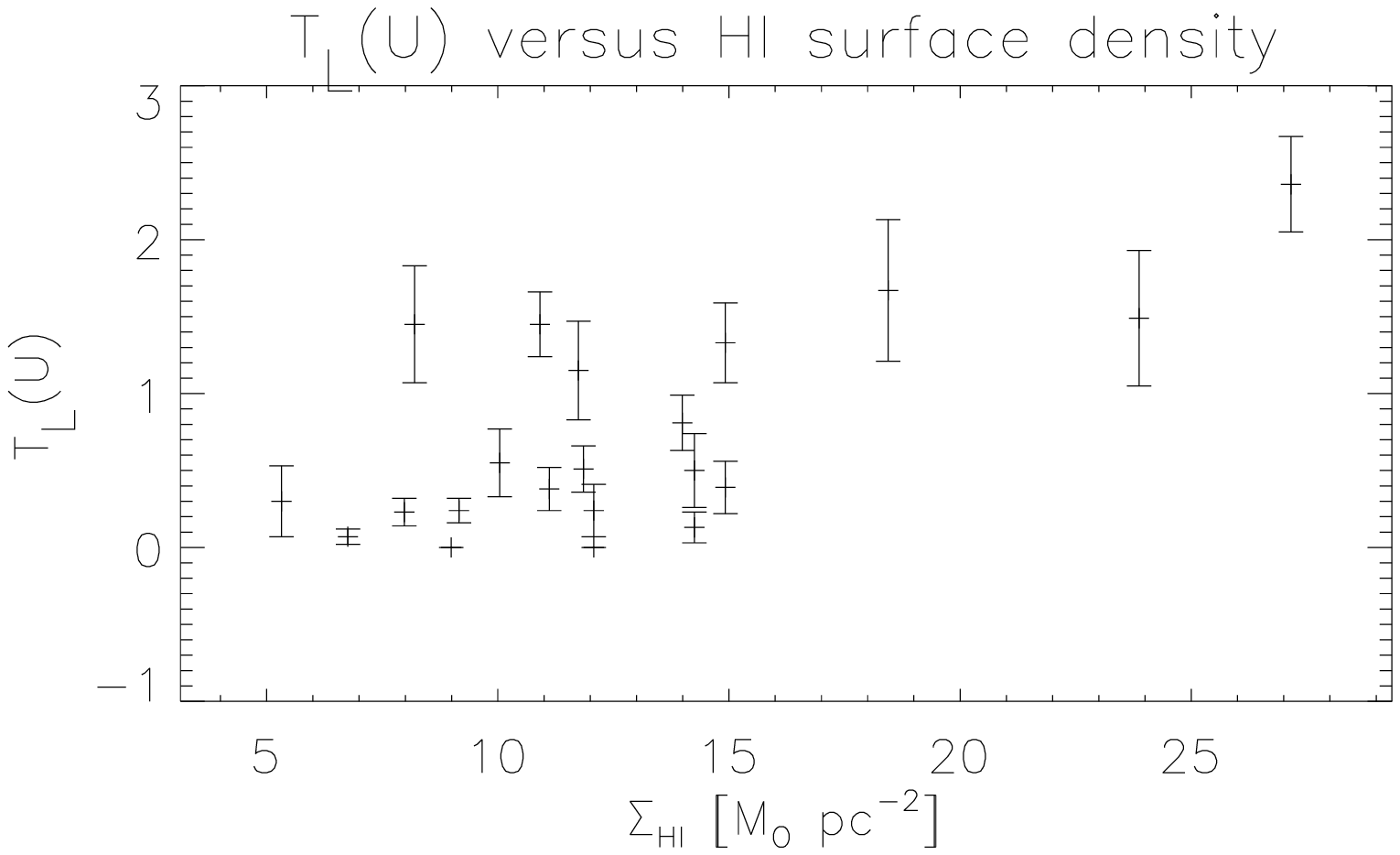}
\caption{\label{fig:sbm21_p1}
  \tlu\ as a function of the HI surface density \SHI .
}
\end{figure}

  The FIR luminosity itself is an indicator of the current SFR through 
heating of dust grains by young stars (Kennicutt \cite{kennicutt98b}). The 
uncertainties on the exact relation are, however, considerable and a 
single calibration is unlikely to apply to all galaxies over a wide range 
in morphological type.  This is mainly because older stellar populations 
also contribute to dust heating, and the ratio of current to past star 
formation varies along the Hubble sequence. Of course, the FIR luminosity 
also suffers the same IMF dependence as any other SFR indicator.
Here we will use the calibration
by Buat \& Xu (\cite{bx96}) which is claimed to be reasonably accurate for
galaxies later than type Sab, noting that it may overestimate the SFR 
in starburst galaxies by about a factor of 2 (Kennicutt \cite{kennicutt98b}):
\begin{equation}
  \mbox{SFR}(\msun \, \mbox{yr}^{-1}) = 
    8^{+8}_{-3} \times 10^{-37} \lfir
  \label{eq:sfr}
\end{equation}
  where \lfir\ is the far-infrared luminosity (in J/sec). We obtain \lfir\
from the \mfir\ magnitudes in Table~\ref{tab:intdata} and the distance
moduli, using the relation
\begin{equation}
  \mfir = -20 - 2.5 \log(S_{\rm FIR})
  \label{eq:mfir_sfir}
\end{equation}
with $S_{\rm FIR}$ being the far-infrared flux density, based on
IRAS $60\mu$ and $100\mu$ flux densities (see RC3 for details). From 
(\ref{eq:sfr}) and (\ref{eq:mfir_sfir}),
\begin{equation}
  \mbox{SFR}(\msun \, \mbox{yr}^{-1}) =
    0.0096^{+0.0096}_{-0.0036} D^2 
     10^{-0.4\times (m_{\rm FIR} + 20)}
\end{equation}
where $D$ is the distance in pc.

However, the global SFR is not likely to tell us much because of the large 
range in galaxy size and total luminosity. We therefore normalise the SFR to 
the area of each galaxy based on the {\it optical} diameter 
(using $\log D_0$ from RC3), defining \Ssfr\ as the SFR per kpc$^2$: 
\begin{equation}
  \Ssfr (\msun \, {\rm yr}^{-1} \, {\rm kpc}^{-2})
    = 144000 \times 10^{-0.4 \, \mfir \, - \, 2 \, \log D_0}
  \label{eq:ssfr}
\end{equation}
It might seem more reasonable to normalise to some area traced by the 
FIR luminosity, but the resolution of the IRAS data does not allow this 
in all cases. Another possibility would be to normalise to the 
optical luminosity rather than the area, but in this way some information
might be lost because the optical luminosity is also correlated with the SFR,
and because of the contribution from the bulge/halo components.

  Fig.~\ref{fig:sfr_p1} displays \tlu\ as a function of the host galaxy
SFR according to Eq.~\ref{eq:sfr} (top panel) and \Ssfr\ (bottom panel).
A correlation is evident in both cases, but the scatter clearly
decreases when plotting \tlu\ as a function of \Ssfr\ rather than
the global SFR.

  The SFR in galaxies is generally assumed to be proportional to some
power of the gas density (Schmidt \cite{schmidt59}), and it has recently
been shown that the Schmidt law can also be formulated in terms of surface
densities with $\Ssfr \propto \Sigma_{\rm gas}^{1.4}$ (Kennicutt 
\cite{kennicutt98a}).  It would therefore be of interest to look for a 
corresponding relation between \tlu\ and the gas surface density 
(\Sgas ). Lacking a homogeneous set of data on total gas masses, we will 
here consider the atomic hydrogen mass $M_{\rm HI}$ which may be derived 
from the 21-cm flux density (Roberts \cite{roberts75}):
\begin{equation}
   M_{\rm HI} (\msun ) \, = \, 2.356 \times 10^{19} D^2 
      \int_{-\infty}^{\infty} S_{\nu} dV_r
   \label{eq:roberts}
\end{equation}
where $D$ is the distance in pc and $\int_{-\infty}^{\infty} S_{\nu} dV_r$ 
is the flux density integrated over the line profile. Here $S_{\nu}$ is in 
units of \mbox{W m$^{-2}$ Hz$^{-1}$} and $V_r$ is in km/sec.  The total 
integrated flux density $S_{\rm HI}$ can be obtained from the $m_{21}$ values 
given in Table~\ref{tab:intdata} using the expression
\begin{equation}
  m_{21} \, = \, 21.6 - 2.5 \log(S_{\rm HI})
  \label{eq:m21}
\end{equation}
with $S_{\rm HI}$ in units of \mbox{$10^{-24}$ W m$^{-2}$} (RC3). 
Combining (\ref{eq:roberts}) and (\ref{eq:m21}) we obtain
\begin{equation}
  M_{\rm HI} (\msun ) \, = \, 4.97 \times 10^{-9} \, D^2 \, 
    10^{0.4 \times (21.6 - m_{21})}
\end{equation}
We ignore corrections for self-absorption since most of the galaxies are
seen nearly face-on. No homogeneous set of data is available on the HI 
sizes so we use again the optical sizes to derive the HI surface
density \SHI :
\begin{equation}
  \SHI (\msun \, {\rm pc}^{-2}) =
     3.26\times10^9 \, \times \, 10^{-0.4 \, m_{21} - 2 \, \log D_0}
  \label{eq:shi}
\end{equation}
This is somewhat problematic since 
HI disks often extend beyond the optical disk size.  However, as long as 
the same procedure is applied to all galaxies in the sample the results 
should at least be comparable, although we stress that the absolute values 
of the HI surface density (\SHI ) should probably not be given 
too much weight.  The uncertainties on $m_{21}$ quoted in RC3 are typically 
of the order of 0.1 mag or about 10\%, so errors in \SHI\ are
more likely to arise from the area normalisation because of
differences in the scale length of the HI disks relative to the optical sizes.

  Fig.~\ref{fig:sbm21_p1} shows \tlu\ vs. \SHI .
The plot clearly shows a correlation, although not as nice as between
\tlu\ and \Ssfr . This may not be surprising, considering the relatively 
small range in \SHI\ compared to \Ssfr , which makes the result much more 
sensitive to errors in the area normalisation. Also, \Ssfr\ (and thus
\tlu ) is expected to
depend on the {\it total} gas surface density \Sgas\ of which \SHI\ 
constitutes only a fraction, which is not necessarily the same from galaxy
to galaxy.
However, we note 
that Kennicutt (\cite{kennicutt98a}) finds that \Ssfr\ correlates nearly as 
well with \SHI\ as with \Sgas\ though the physical interpretation of the
correlation between \Ssfr\ and \SHI\ is not entirely clear, because 
of the complicated interplay between the different phases of the interstellar 
medium and young stars.  Somewhat surprisingly, Kennicutt 
(\cite{kennicutt98a}) finds no 
significant correlation between \Ssfr\ and the surface density of 
{\it molecular} gas.  

\subsection{Including literature data}

  It is of interest to see if the \tlu\ vs. \Ssfr\ relation holds also
when including other types of galaxies than those from Paper1. In 
particular, a comparison with the many studies of starburst galaxies that
exist in the literature is tempting. In Tables~\ref{tab:bprop} and
\ref{tab:intdata} we have included literature data for a number of different
galaxies, briefly discussed in the following. These galaxies have been
chosen mainly so that a number of different cluster-forming environments
are represented, with the additional criterion that some photometry
was available for individual clusters so that (at least approximate)
\tlu\ values could be estimated.

  We first give a few comments on each galaxy:

\paragraph{\object{NGC 5253}:}

  A dwarf galaxy, located at a projected distance of about 130 
kpc from NGC~5236. It is possible that the starburst currently going on in
this galaxy could have been triggered by interaction with its
larger neighbour, though no obvious indications of direct interaction
between the two galaxies are evident. Several massive clusters exist
in NGC~5253, but the absolute magnitudes are somewhat uncertain because
of heavy extinction (Gorjian \cite{gorjian96}). 

\paragraph{\object{NGC 1569} and \object{NGC 1705}:}

  These were two of the first galaxies in which the existence of
``super star clusters'' was suspected (Arp \& Sandage \cite{as85}).
Their \tlu\ values are dominated by 2 bright clusters in NGC~1569 and
by a single cluster in NGC~1705, each with $M_V \approx -13$. 
Both galaxies are gas-rich amorphous dwarfs, but none of them
have high enough star formation rates to qualify as real starburst galaxies
(O'Connell et al. \cite{oconnell94}) although NGC~1569 may be in
a post-starburst phase (Waller \cite{waller91}).

\paragraph{\object{NGC 1741}:}

  A merger/starburst galaxy with a large number of very young ($\sim 10$ Myr)
YMCs. Johnson et al. (\cite{johnson99}) found that YMCs contribute with 5.1\% 
of the $B$-band luminosity in NGC~1741, and since the YMCs are generally bluer 
than the host galaxy we have crudely adopted \tlu\ $\sim 10$ for 
Table~\ref{tab:bprop}.

\paragraph{\object{NGC 1275}:}

  This is the central galaxy in the Perseus cluster. It is sitting at
the centre of a cooling flow, and exhibits a number of structural
peculiarities (N{\o}rgaard-Nielsen et al. \cite{norgaard93}). Most
recently, the cluster system in NGC~1275 was studied by Carlson et al.
(\cite{carlson98}) who identified a population of 1180 YMCs. It has
been proposed that the clusters could have condensed out of the cooling
flow, but it seems more likely that they are due to a merger
event (Holtzman et al. \cite{holtzman92}).

\paragraph{\object{NGC 3256}:}

  This is one of the classical recent merger galaxies. Zepf et al. 
(\cite{zepf99}) identified more than 1000 YMCs on HST / WFPC2 images, and 
estimated that the
clusters contribute with about 15--20\% of the total $B$-band luminosity in
the starburst region. Thus, we adopt \tlu\ = 15.

\paragraph{\object{NGC 3921}:}

  NGC~3921 is the remnant of two disk galaxies which merged 
0.7$\pm$0.3 Gyr ago, and contains about 100 YMC candidates with $V-I$
colours consistent with this age (Schweizer et al. \cite{schweizer96}). We 
have calculated \tlu\ using the objects classified as types 1 or 2 by 
Schweizer et al. (\cite{schweizer96}).

\paragraph{\object{NGC 7252}:}

  Another famous example of a merger galaxy, although 
dynamically more evolved than NGC~3256 and the Antennae. The merger age
has been estimated to be about 1 Gyr (Schweizer \cite{schweizer82}),
and the 140 YMCs that have been identified in the galaxy have colours
roughly compatible with this age (Whitmore et al. \cite{whitmore93};
Miller et al. \cite{miller97}).

\paragraph{\object{IC 1613}:}

  IC~1613 stands out by containing very few star clusters at all, even when
counting ``normal'' open clusters (van den Bergh \cite{vanden79}).
Indeed, it has the lowest star formation rate among all the galaxies discussed
in this paper and thus fits nicely into the \tlu\ vs. SFR relation.

\paragraph{} 

\begin{figure}
\begin{minipage}{88mm}
\epsfxsize=88mm
\epsfbox[87 368 555 640]{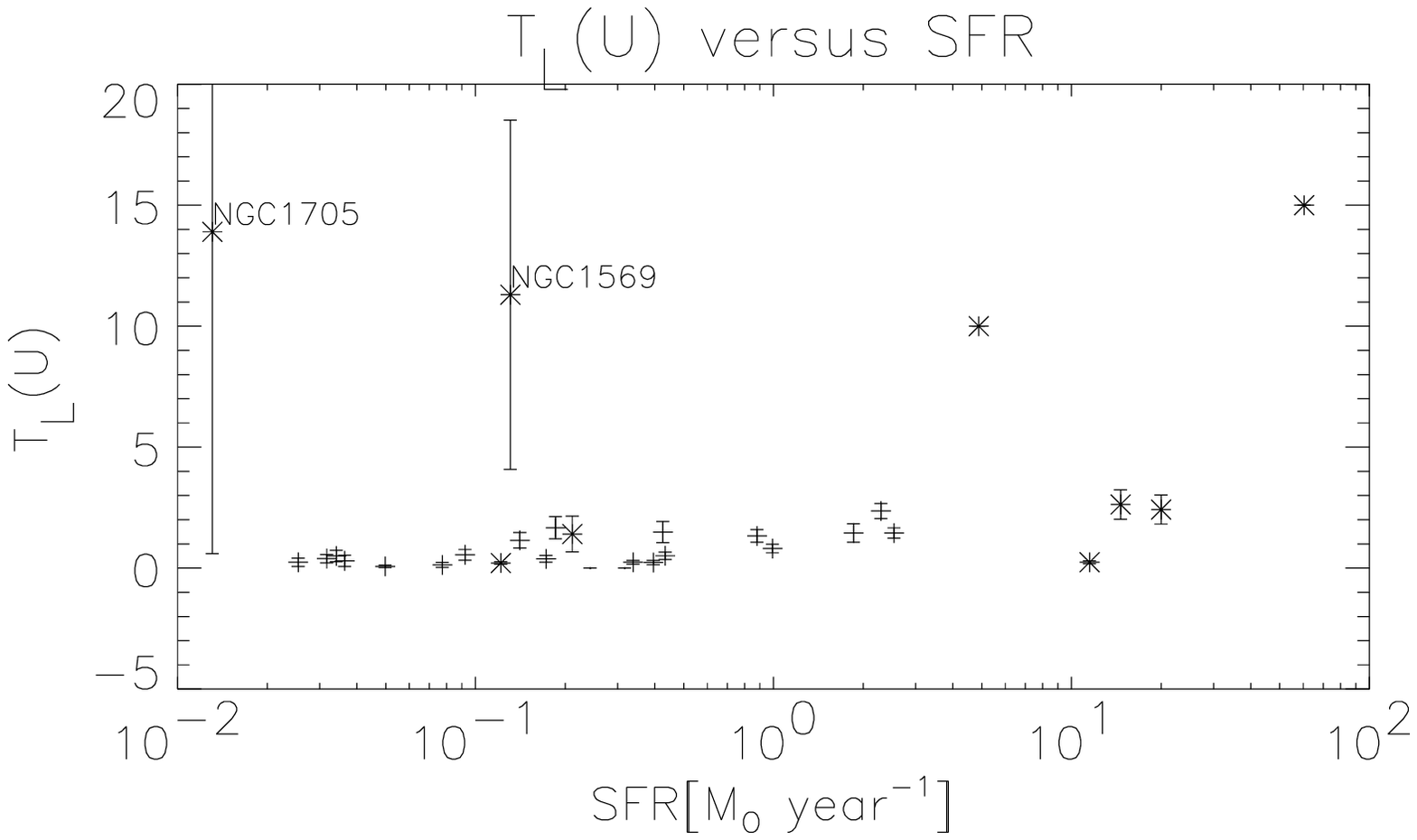}
\end{minipage}
\begin{minipage}{88mm}
\epsfxsize=88mm
\epsfbox[87 368 555 640]{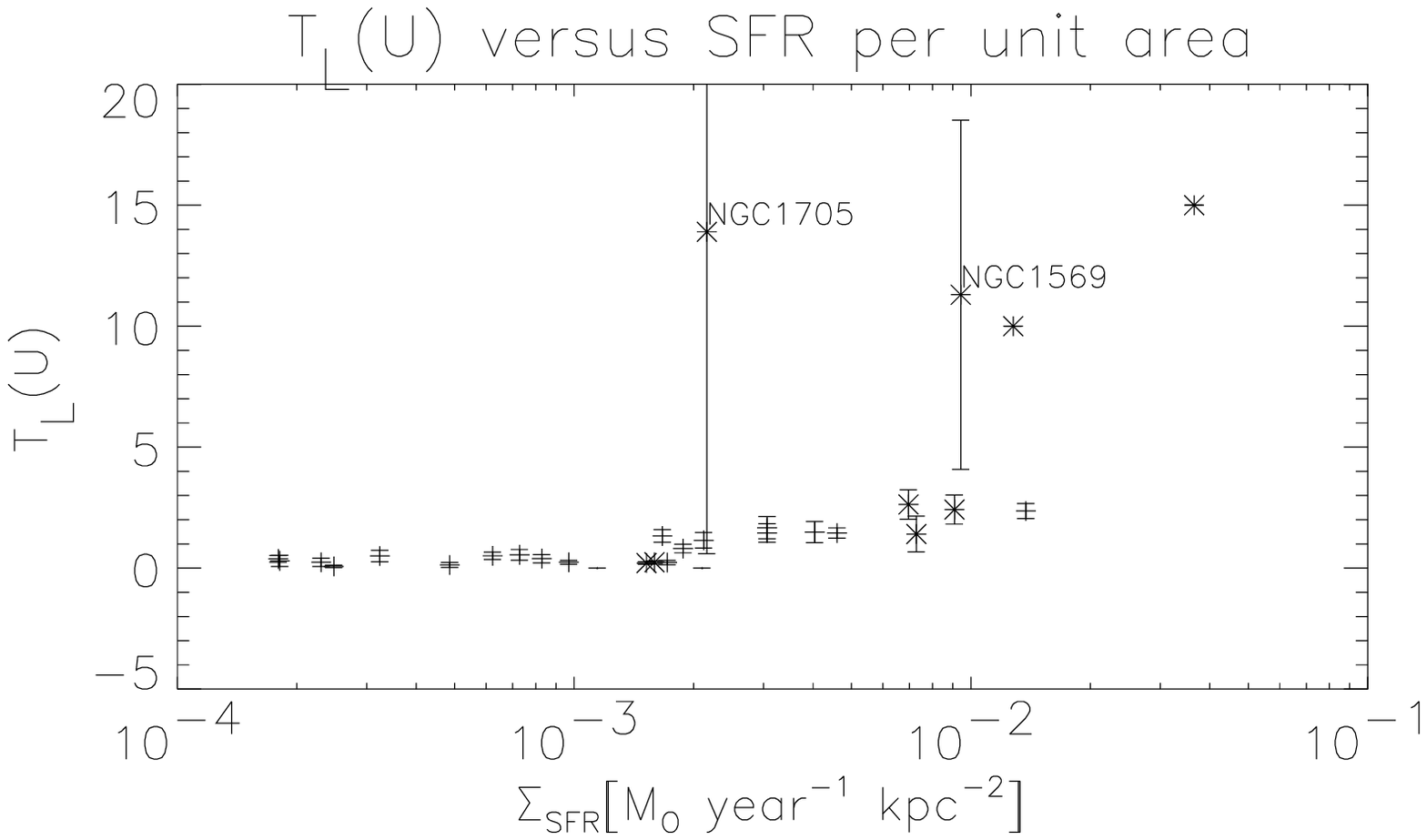}
\end{minipage}
\caption{\label{fig:sfr}
  \tlu\ vs. Star Formation Rate and \Ssfr\ as derived from the FIR 
luminosities for 
all galaxies in Table~\ref{tab:bprop}. Data from Paper1 are shown with $+$ 
markers while literature data are plotted with $*$ markers.
See caption to Fig.~\ref{fig:sfr_p1} for further details.
}
\end{figure}

\begin{figure}
\epsfxsize=88mm
\epsfbox[60 370 550 640]{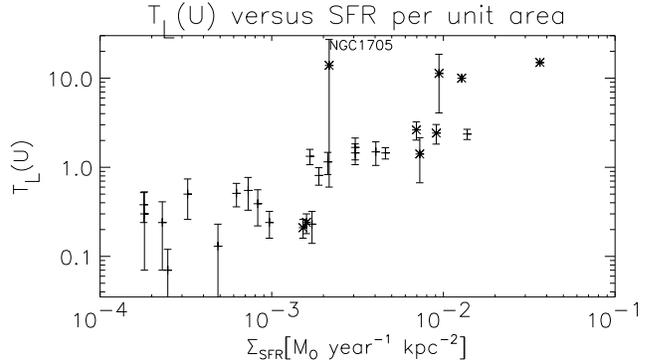}
\caption{\label{fig:sfr_log}
  Same as Fig.~\ref{fig:sfr}, lower panel, but with logarithmic $y$-axis.
Galaxies with $\tlu = 0$ have arbitrarily been assigned
$\tlu = 10^{-3}$.
}
\end{figure}

\begin{figure}
\epsfxsize=88mm
\epsfbox[87 368 555 640]{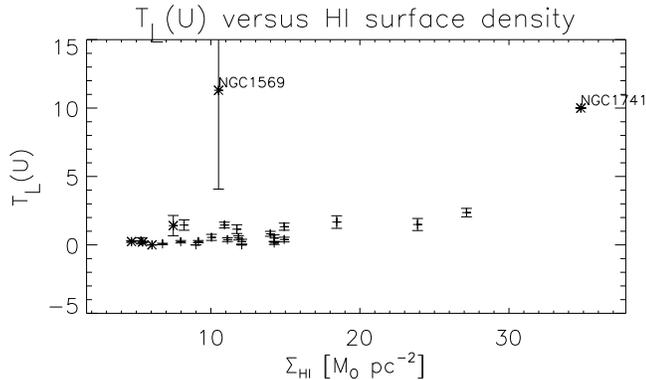}
\caption{\label{fig:sbm21}
  \tlu\ as a function of neutral hydrogen surface density \SHI\ for all
galaxies with 21-cm data.
}
\end{figure}

\begin{figure}
\epsfxsize=88mm
\epsfbox[87 368 555 640]{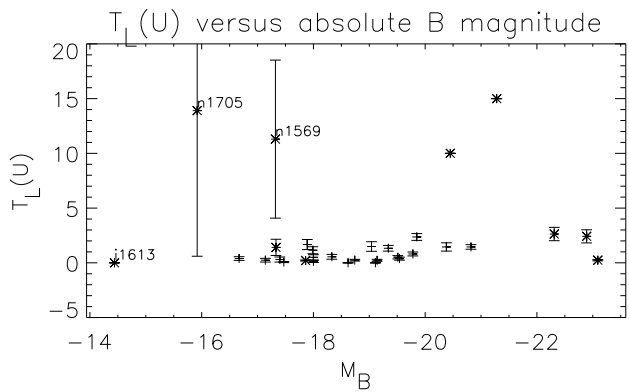}
\caption{\label{fig:mb_tlu}
  \tlu\ as a function of the absolute host galaxy B magnitude.
}
\end{figure}

  The conclusion that \Ssfr\ may be one of the dominating parameters in 
determining the properties of the young cluster systems in galaxies is 
further strengthened by including the literature data for a variety of star 
forming environments.  Fig.~\ref{fig:sfr} shows \tlu\ as a function of 
the global SFR and \Ssfr\ once again, but now with all galaxies in 
Table~\ref{tab:bprop} included. \tlu\ now ranges from 0 -- 15, and the 
galaxies span 5 decades in global SFR. Like in Fig.~\ref{fig:sfr_p1}, \tlu\ 
correlates significantly better with \Ssfr\ than with the global SFR.
The two dwarf galaxies NGC~1569 and NGC~1705, especially the 
latter, deviate somewhat from the general pattern, but because the cluster
light in both these galaxies is dominated by only a few bright clusters,
the statistical significance of their high \tlu\ values is low. Furthermore, 
the area normalisation is obviously uncertain and could easily shift the data 
points horizontally in the diagram by large amounts.   The 
data presented here are compatible with a linear relation between \Ssfr\ 
and \tlu , though a least-squares fit formally yields a power-law dependence 
of the form $\tlu \sim \Ssfr^{0.87\pm0.15}$. This is seen somewhat more 
clearly on a double-logarithmic plot (Fig.~\ref{fig:sfr_log}).

  The \tlu\ vs. \SHI\ diagram for all galaxies with 21 cm data in RC3 is 
shown in Fig.~\ref{fig:sbm21}. Note that $m_{21}$ data are lacking for many 
of the starburst and merger galaxies in Table~\ref{tab:intdata}.  Thus, 
the only galaxies in Fig.~\ref{fig:sbm21} with a significantly higher 
\tlu\ value than those from the Paper1 sample are NGC~1569 and NGC~1741. 
Again we see the poor fit of NGC~1569 into an otherwise quite good 
correlation, while NGC~1741 is located to the far right in the diagram, 
as expected from its high \tlu\ value.  

  NGC~1569 and NGC~1705 differ from the other cluster-rich galaxies by
their relatively low absolute luminosities, and one could speculate that
YMC formation might be due to a different physical mechanism in these 
galaxies. In Fig.~\ref{fig:mb_tlu} we show \tlu\ as a function of the 
absolute $B$ magnitude of the host galaxy (derived from $m_B$ and the 
distance moduli and $A_B$ values in Table~\ref{tab:bprop}). Although
NGC~1569 and NGC~1705 are among the least luminous galaxies in our
sample, there are in fact even less luminous galaxies with ongoing
star formation, but without rich cluster populations (notably IC~1613).
Thus the main cause for the high \tlu\ values of NGC~1705 and NGC~1569 
still appears to be their relatively high level of star formation 
activity, and the poor fit of these two galaxies into the
\tlu\ -- \Ssfr\ relation may be ascribed primarily to the small number
statistics of their cluster systems.

\section{Discussion}

  Our data apparently indicate that the formation efficiency of YMCs in 
galaxies is closely linked to the star formation activity.  By using $U$-band 
luminosities, the derived specific luminosities are dominated by the youngest 
stars, effectively making \tlu\ a measure of the relative fraction of stars 
that currently form in massive clusters. \tlu\ increases from about 
0.1 in the most cluster-poor galaxies to 15 or more in merger galaxies
like NGC~3256. We can, of course, not exclude the possibility that some of 
the very youngest objects are unbound associations that will not survive 
for long, rather than bound clusters. However, as shown in Paper1,  the
age distributions of the clusters are generally quite smooth, indicating
that at least some fraction of the objects are indeed gravitationally
bound star clusters, orders of magnitudes older than their crossing
times (Larsen \cite{larsen99}).

  The \tlu\ vs. \Ssfr\ correlation may explain why YMCs have, so far, been
noticed predominantly in late-type galaxies (Kennicutt \& Chu \cite{kc88}).
Apart from the small number of nearby, early-type spirals, this may just
be an effect of the general increase in SFR along the Hubble sequence. 
However, there is a large scatter in SFR at any given morphological type 
(Kennicutt \cite{kennicutt98b}), which is presumably also the reason for 
the corresponding scatter in \tlu , and we would expect YMCs to be abundant 
also in Sa and Sb galaxies with a sufficiently high \Ssfr\ (that is, higher 
than about $10^{-3}$ \msun\ yr$^{-1}$ kpc$^{-2}$, Fig.~\ref{fig:sfr_p1}). 
The main point here is that \tlu\ correlates with the SFR, rather than that 
formation of YMCs is generally favoured in late-type galaxies. 
  
  Our data imply a continuum of \tlu\ values, varying smoothly with \Ssfr ,
rather than a division of galaxies into those that contain YMCs and those
that do not. That YMCs have often been considered as a special class of 
objects which only exist in certain galaxies, probably arises from the fact 
that most efforts to detect them have focused on starburst galaxies, where 
they are much more numerous. Table~\ref{tab:bprop} also shows that the $M_V$ 
of the brightest cluster in each galaxy varies significantly. Recent, deep 
studies of young clusters 
in NGC~3256 (Zepf et al. \cite{zepf99}), NGC~1275 (Carlson et al. 
\cite{carlson98}) and other galaxies have not revealed any clear indications 
of a turn-over in the cluster luminosity function down to $M_V \sim -8.5$ or 
so, so the fact that these galaxies contain brighter clusters than less 
cluster-rich systems may just be a statistical effect.

There does not seem to be any SFR threshold for formation of YMCs. Instead,
the number of YMCs formed and the efficiency of YMC formation appear to
increase steadily with the star formation rate. This also raises the question 
whether massive star clusters are good tracers of the star formation history 
in a galaxy, as they have often been used in the Magellanic Clouds.
For example, the apparent lack of massive star clusters in the LMC in the
age range 4 -- 10 Gyr (Girardi et al. \cite{girardi95}) has been seen as an 
indication that the LMC was in a sort of ``hibernating'' state during this 
period. However, if the cluster formation efficiency depends upon the star 
formation rate as suggested by this paper, then the ``gap'' in the LMC 
cluster age distribution could merely represent an epoch where star formation 
proceeded at a somewhat slower, but not necessarily vanishing rate. Indeed, 
this has been recently demonstrated from field star studies by Dirsch et 
al. (\cite{dirsch99}).

  It still remains to be explained {\it why} the formation of YMCs is 
correlated with the star formation rate. It is not even clear if YMCs form 
because there is a high SFR, or if the \tlu\ -- \Ssfr\ correlation is 
a consequence of some underlying mechanism that regulates both the SFR and 
the formation of YMCs. Here we briefly discuss both  possibilities in a 
speculative manner, and consider how they may complement each other.

\subsection{SFR and cluster formation as resulting from a high gas density}

  An underlying parameter controlling both the star 
formation rate and the ability to form bound, massive clusters could be 
the {\it mean} gas density. It is well established that the SFR in a galaxy 
scales with some power of the gas density. Denoting the total gas surface 
density \Sgas , the Schmidt (\cite{schmidt59}) law may be written as 
$\Ssfr \propto \Sigma_{\rm gas}^N$, where the exponent $N$ has a value close 
to 1.4 (Kennicutt \kena ). As shown by Kennicutt (\cite{kennicutt98a}), the 
Schmidt law provides a surprisingly good description of the SFR in galaxies 
in terms of a global \Sgas\ over a wide range of surface gas density, so 
there is hope that cluster formation may depend on similar global galaxy 
properties, at least to a first approximation.

  The \tlu\ -- \Ssfr\ relation in combination with the Schmidt law implies
that \tlu\ should scale with \Sgas\ as well. This is at least partly confirmed
by the observed correlation between \tlu\ and the HI gas surface density, 
\SHI\ (Figs.~\ref{fig:sbm21_p1} and \ref{fig:sbm21}). A \tlu\ -- \Sgas\ 
relation may 
follow from the fact that a higher gas density leads to a generally higher 
ISM pressure ($P \sim \Sigma_{\rm gas}^2$ where $P$ is the pressure, Elmegreen
\cite{elmegreen99}). The ISM pressure has been suggested to be one of 
the dominant parameters governing the formation of strongly bound clusters 
(Elmegreen \& Efremov \cite{ee97}) and acts by producing proto-cluster 
clouds with higher binding energies, thus preventing them from dispersing 
too easily once star formation sets in. The clouds will have higher 
densities so that recombination rates are higher, and smaller fractions of 
the gas will be ionized by massive stars. Also the dispersive power of stellar
winds and supernovae will be lower in a high density environment. All these
effects promote a high star formation efficiency, one of the necessary
conditions to produce a bound cluster.

  A \tlu\ -- \Sgas\ relation may thus be explained by saying that 
the high gas density delivers the required high pressure to form massive 
clusters. As local fluctuations are always important, we do not expect an 
overall ``threshold'' gas density when averaging over a whole galaxy, but 
as \Sgas\ increases, the number of regions with the required high density 
will gradually increase too and naturally lead to the formation of more 
strongly bound clusters. With a high \Sgas\, one also expects a fast growth 
of the protocluster so that higher masses become plausible.

\subsection{A high SFR as a precondition to form massive clusters}

The main effect of a high SFR is to pump energy into the ISM. Can this 
energy be responsible of creating suitable conditions for globular clusters?
According to Elmegreen \& Efremov (\cite{ee97}), globular cluster formation
needs  highly efficient star formation in a high pressure environment.   

  In order to form a massive, bound cluster two timescales apparently 
are of importance: The timescale for formation of a cloud core, which is 
massive enough to host a massive cluster, \tcc , and the time scale for 
(high-mass) star formation in the cloud core, \tsf.

  It is interesting to note that the {\it average} density of a
proto-YMC cloud prior to the onset of star formation  (if the radius of
the cluster equals the radius of the proto-cluster cloud)
\begin{equation}
  \rho \approx 1.3 \times 10^{-20} \left(\frac{M}{10^5 \, \msun}\right)
      \left(\frac{R}{5 \, {\rm pc}}\right)^{-3} \, {\rm g} \, {\rm cm}^{-3}
\end{equation}
must be quite similar to that observed in cluster-forming clumps in Galactic
giant molecular clouds (Lada et al. \cite{lada97}), although the total
mass is much larger. In the Milky Way, efficient cluster formation appears
to take place only in massive, high-density cloud cores, but not in
{\it all} such cores (Lada et al. \cite{lada97}). A discriminating factor
appears to be the degree of fragmentation within the core, presumably because
star formation takes place only in regions with a density higher than
$10^5$ molecules per cm$^3$, or about $3 \times 10^{-19} \,
{\rm g} \, {\rm cm}^{-3}$.
If such a critical density exists, one could understand \tsf\ as the timescale
which is needed for the gas to reach this density. 

 Whatever the 
formation mechanism of the cloud core is (Elmegreen \cite{elmegreen93}),
star formation may not commence early, because the returned energy from 
massive stars by radiation, outflows and stellar winds 
presumably will terminate the growth of the cloud core and moreover is 
a threat to its dynamic stability. If $\tsf \ll \tcc $, the result might be 
a low mass cluster. 

In addition, $\tsf$ may not vary strongly
in the cloud core. If it did so, one expects the outcome again to be not a globular
cluster, but a star forming region with many dynamically 
distinct smaller clusters. i.e. a configuration resembling an association. 
However, if $\tsf > \tcc $, the cloud core can grow undisturbed by star 
formation and develop towards a strongly bound state. This may be the case
either if the onset of star formation is somehow delayed, or if the
formation of the proto-cluster cloud proceeds rapidly.

Any attempt to construct a scenario is hampered by  the fact that even the 
physical cause for the onset of star formation
(e.g. ambipolar diffusion, Jeans instabilities, thermal instabilities) 
is not yet clearly identified. However, star formation in general means to
put matter into a state of strongly negative potential energy, so there 
is demand for an external energy input to delay star formation, even if the 
exact process is not known. 
 

Part of the required energy may come from early 
low-mass star formation within the cloud (Tan \cite{tan99}), but in order 
to maintain energy equilibrium in a large, massive cloud, external heat 
sources might also be necessary. At the highest densities 
($\ga 10^5 \, {\rm cm}^{-3}$) the thermal pressure may become able to
compete with or even dominate over magnetic pressure 
(Pringle \cite{pringle89}), so an energy input may also prevent premature 
star formation by Jeans or thermal instability (Murray \& Lin \cite{ml89}).


  A high overall star formation rate naturally provides a number of energy 
sources, not only in the form of radiation from massive stars. Other 
possibilities are supersonic motions in the gas, induced by supernova
shells or stellar winds. These may also help to compress proto-cluster 
clouds, so that large amounts of gas can be collected at high densities 
more easily, and fast enough to form a bound cluster.  There is, in fact, 
some evidence that the formation of massive clusters marks the culmination 
of episodes of vivid star formation (Larson \cite{larson93}). 

  These arguments apply not exclusively to massive clusters, but it is 
conceivable that more extreme external conditions can lead to denser, more 
massive clusters. This is in good agreement with the observed continuous 
dependence of \tlu\ on \Ssfr.

\subsection{The relation to old globular clusters}

  Within the scenarios described above, some findings regarding the 
systematics of globular clusters in early-type galaxies become 
understandable. The relevant labels can be called ``hot'' and ``cold'' 
dynamical environments. Cluster formation in orderly rotating gaseous disks, 
a ``cold'' dynamical environment, may not be supported without the impact 
of a high star formation rate. In the dynamically ``hot'' bulges and halos, 
the external energy supply comes from turbulent motions in the ambient 
medium which acts as a reservoir.

  A striking feature regarding cluster populations in elliptical galaxies 
is the high specific frequency of central galaxies in clusters like M87 
and NGC 1399. At least in the case of NGC~1399, these can be understood by the 
early infall of a population of dwarf galaxies into the Fornax cluster
(Hilker et al. \cite{hilker99}). The infall velocities are of the order 
hundreds of km/s and the kinematic situation is similar to those in 
starburst galaxies. A lot of energy can by dissipated and very suitable 
conditions for cluster formation are provided.  

The same interpretation may be valid for the relation between the specific
frequency of globular cluster systems and the environmental galaxy density 
of the host galaxies (West \cite{west93}): The higher the galaxy density, the 
more frequent galaxy interaction with violent star formation must have been,
leading to higher cluster formation efficiencies.
   
  This might have been generally the case  in the very early Universe, when 
the average star formation rates were much higher than nowadays. The old halo 
globular cluster systems of ``normal'' galaxies, which belong to the oldest
stellar populations in galaxies, have been formed during this period, which
quite naturally provided suitable conditions for massive cluster formation.

\section{Conclusions}

  We have studied the cluster systems of the 21 galaxies in the sample 
of Larsen \& Richtler (\cite{lr99}) together with literature data for some
additional galaxies.  It has been demonstrated that the specific $U$-band
luminosity of the cluster systems, \tlu\ (Eq.~(\ref{eq:tu})) correlates with
host galaxy parameters indicative of the star formation rate, in particular
the \mbox{$B$-band} surface brightness ($m_{25}$) and IRAS far-infrared
fluxes. Using the FIR fluxes to derive star formation rates (SFR) and
obtaining the area-normalised SFR (\Ssfr ), we find an even stronger
correlation with \tlu , which seems to indicate that the formation of YMCs is 
favoured in environments with active star formation.  However, this does not 
imply that YMCs form only in bona-fide starbursts, but rather that the cluster 
formation efficiency as measured by \tlu\ increases steadily with \Ssfr\
and that the formation of YMCs in starbursts and mergers may just be 
extreme cases of a more general phenomenon.

  We have also compared the \tlu\ values with integrated HI gas surface
densities (\SHI ) and find a correlation here as well. Since \tlu\ and 
\Ssfr\ are correlated, this is an expected consequence of the fact that 
\Ssfr\ scales with some power of the gas surface density \Sgas\
(Kennicutt \kena ). 

  Although the two amorphous dwarfs \object{NGC~1569} and \object{NGC~1705}
have rather high \tlu\ values for their star formation rates, we do not see 
any examples of \emph{cluster-poor} galaxies with a \emph{high} \Ssfr . In 
other words, a galaxy contains large numbers of YMCs whenever \Ssfr\ is high 
enough, although the physical relation is not yet well understood.
Formation of a rich cluster system does not 
require a strong spiral density wave, for example, since the flocculent 
galaxy \object{NGC~7793} has a high \tlu . Interaction with nearby neighbours 
does not appear to be necessary either, as illustrated by \object{NGC~1156} 
which has been labeled ``the less disturbed galaxy in the Local Universe'' 
(Karachentsev et al.\ \cite{kara96}), but nevertheless contains a rich 
population of YMCs.

  Some mechanisms were outlined which may explain why massive star
clusters form at a high efficiency in environments with a high SFR:
A generally high SFR acts as an energy source that keeps molecular clouds in 
an equilibrium state and allows massive clouds to contract to a high
density before high-mass star formation sets in. Once the required high 
average density to form a YMC is reached (about $10^4$ cm$^{-3}$), star 
formation proceeds rapidly and at a high efficiency within the clouds, 
because the high pressure in the ambient medium keeps the proto-cluster 
clouds from dispersing (Elmegreen \& Efremov \cite{ee97}).

\begin{acknowledgements}
  This research was supported by the Danish Natural Science Research Council
through its Centre for Ground-Based Observational Astronomy.
This research has made use of the NASA/IPAC Extragalactic Database (NED)
which is operated by the Jet Propulsion Laboratory, California Institute
of Technology, under contract with the National Aeronautics and Space
Administration. We would like to thank B. Elmegreen for interesting 
discussions.
\end{acknowledgements}

\end{document}